\documentclass[twocolumn]{aastex631}

\usepackage{bbold}
\usepackage{hyperref}
\usepackage{tabularx} 
\usepackage{mathtools}

\hypersetup{
		colorlinks=true,       		
    	linkcolor=blue,          	
    	citecolor=purple,        		
    	filecolor=magenta,      		
		urlcolor=blue,
		bookmarksdepth=4
}

\newcommand{\healpix}[1]{\texttt{HEALPix} #1}

\newcommand{\lz}{\textbf{LOW-Z}}
\newcommand{\hz}{\textbf{HIGH-Z}}

\begin{document}
\title{Investigating cross-correlations between cosmic microwave background lensing and 21 cm intensity mapping}

\author[0000-0002-6519-6038]{Alessandro Marins}

\affiliation{Department of Astronomy,
University of Science and Technology of China, Hefei 230026, China}
\affiliation{School of Astronomy and Space Science, University of Science and Technology of China, Hefei 230026, China}

\author[0000-0001-7438-5896]{Chang Feng}
\altaffiliation{Corresponding author: changfeng@ustc.edu.cn}
\affiliation{Department of Astronomy,
University of Science and Technology of China, Hefei 230026, China}
\affiliation{School of Astronomy and Space Science, University of Science and Technology of China, Hefei 230026, China}

\author[0000-0003-2063-4345]{Filipe B. Abdalla}
\affiliation{Department of Astronomy,
University of Science and Technology of China, Hefei 230026, China}
\affiliation{School of Astronomy and Space Science, University of Science and Technology of China, Hefei 230026, China}

\begin{abstract}
The neutral hydrogen (HI) signal is a crucial probe for astrophysics and cosmology, but it is quite challenging to measure from raw data because of bright foreground contaminants at radio wavelengths. Cross-correlating the radio observations with large-scale structure tracers (LSS) could detect faint cosmological signals since they are not correlated with the foreground, but exquisite component separation procedures must be performed to reduce the variance induced by the foreground. In this work, we adopt the lensing of the cosmic microwave background (CMB) as the LSS tracer and investigate the cross-correlation of CMB lensing and HI observations at the post-reionization epoch. We use simulations to study lensing and HI cross-correlations in the context of next-generation CMB and intensity mapping experiments. We investigate the impact of the component separation based on linear combinations of the HI observations at different frequencies and estimate the signal-to-noise ratios for the cross-correlation measurements in different scenarios. 
\end{abstract}

\section{Introduction}

The 21 cm emission of neutral hydrogen (HI) can trace the evolution of our universe from the dark ages to the post-reionization epochs and is a crucial probe for astrophysics and cosmology. However, the radio foreground is a few orders of magnitude brighter than the HI signal at radio wavelengths. By cross-correlating observations at radio frequencies with large-scale structure (LSS) tracers, such as galaxies, evidence for a 21 cm signal has been found \citep{chang2010, masui2013, anderson2018, wolz2022}. A recent HI detection was also performed through cross-correlation with a galaxy survey where the HI observation was taken from the MeerKAT telescope \citep{santos2005} for several hours of observation and over an effective survey area of approximately 200 deg$^2$, as reported in \cite{cunnington2023}. The cross-correlation between HI intensity mapping (IM) and LSS tracers has the benefit of easily mitigating different systematics and will be complementary to recent auto-power spectrum measurements~\citep{2023arXiv230111943P, 2023ApJ...954..139L, 2024arXiv240800268P}.

The cosmic microwave background (CMB) lensing is a crucial LSS tracer of the underlying dark matter distribution. It can map the accumulated deflections of CMB photons from the last scattering surface to the observers in different sky directions. Moreover, CMB lensing is a unique probe in that it is less affected by astrophysics and non-linearities. A cross-correlation between HI IM and CMB lensing could both confirm the detection of the 21 cm signal and provide complementary cosmological information. However, previous studies revealed particular challenges regarding cross-correlations \citep{karacayli2019, modi2019, moodley2023}. Foreground contaminants are expected to dominate the lowest radial modes in Fourier space, and several methods have been proposed to recover these modes without considering HI signal suppression after blind foreground removal algorithms \citep{karacayli2019, modi2021, carol2022, moodley2023}.


Cosmic shear measurements can be obtained by measuring shapes of background galaxies that are distorted by intervening dark matter \citep{bartelmann2001}. Similar to CMB lensing, the cosmic shear is also correlated with the HI. As demonstrated in \cite{Anut2024}, the cross-correlation of the HI IM from the mid-frequency of the Square Kilometer Array (SKA-MID) and MeerKAT and cosmic shear from optical surveys could also be detected after precise foreground removal of the HI IM observations.

\begin{figure*}[htb!]
    \includegraphics[scale=0.335]{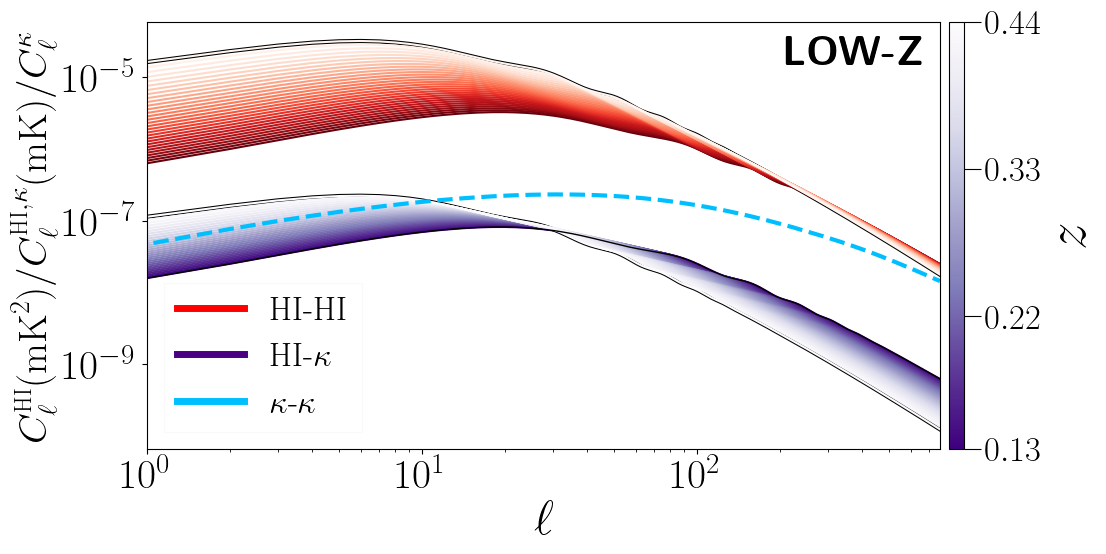}
    \includegraphics[scale=0.335]{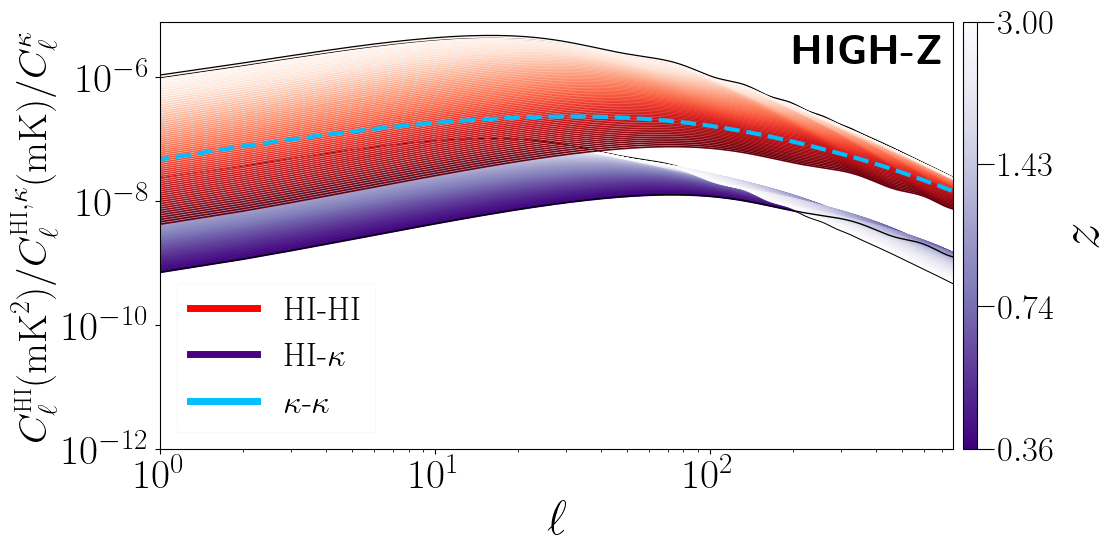}
    \caption{The red gradient colors describe the theoretical HI power spectra in unit of $\textrm{mK}^2$, the blue dashed line is the theoretical CMB lensing power spectrum $\kappa$-$\kappa$, and the purple gradient colors correspond to the theoretical HI-$\kappa$ cross-correlations in unit of $\textrm{mK}$. The gradient evolution changes from darker colors for the lower redshifts to whiter colors for the higher redshifts. In this work, we consider two representative scenarios for the HI IM: \lz\, corresponding to HI observations from 980 to 1260 MHz (0.13$<z<$0.45), and \hz\, from 350-1050 MHz (0.39$<z<$3.05).}
    \label{fig: theory}
\end{figure*}

In this work, we use end-to-end simulations to study complex cross-correlations between the HI and CMB lensing, incorporating blind foreground removal procedures, such as principal component analysis (PCA) \citep{ShifanZuo2023}, independent component analysis (ICA) \citep{maino2002}, general morphological component analysis (GMCA) \citep{bobin2007}, and generalized needlet internal linear combination (GNILC) methods \citep{GNILCplanck2016}. In this work, we adopt the FastICA method as a benchmark test.

The organization of this work is as follows: we describe the theoretical models for HI and CMB lensing in Sec. 2 and the simulations and validations in Sec. 3. We describe the radio sky model and study the foreground removal procedures in Sec. 4. We present cross-correlation results in Sec. 5 and discuss simulations with experimental specifications for HI IM and CMB lensing in Sec. 6. We conclude in Sec. 7. \\

\section{Theoretical models for the cross-correlation signals}

We define the three-dimensional dark matter distribution as $\delta(\textbf{x})=\delta(\chi\textbf{n})$, where $\chi$ is the comoving distance, $\textbf{n}$ is a direction in the sky, and $\textbf{x}$ denotes the real space coordinates. The CMB lensing convergence field $\kappa$ is a projected field, which is a sum of the weighted density field as 
\begin{eqnarray}
    \kappa (\textbf{n}) &=& \int \textrm{d}\chi W_{\kappa}(\chi) \delta (\chi\textbf{n}),
\end{eqnarray}
where the weighting function is 
\begin{eqnarray}
    W_{\kappa} \left(z\right) &=& \frac{3\Omega_{m}H_{0}^2}{2 c H(z)} \frac{\chi(z)}{a(z)}\frac{\chi(z_*) - \chi(z)}{\chi(z_*)}.
    \label{eqn: kappa kernel}
\end{eqnarray}
Here, $\Omega_{m}$ is the fraction of matter-energy density, $H(z)$ is the Hubble constant at redshift $z$, $H_{0}$ is the Hubble constant today, $c$ is the speed of light, $a$ is the scale factor, which is $1/(1+z)$, and $z_{\ast}$ refers to the redshift at the last scattering surface.

The HI brightness temperature field is a biased tracer of the underlying matter distribution, so it can be expressed as
\begin{eqnarray}
    \delta  T_{\textrm{HI}}(\textbf{n},\nu_j) = \int \textrm{d}z b_{\textrm{HI}}(z) \bar{T}_{\textrm{HI}}(z)W_{\Delta_j} (z) \delta (\chi\textbf{n})\label{brightT},
\end{eqnarray}
with a bias $b_{\textrm{HI}}(z)$. $W_{\Delta_j} (z)$ is the window function derived from the radio bandwidth $\Delta_{\tiny j}$, which is assumed to be a top-hat window. The brightness temperature described in Eq. (\ref{brightT}) is essentially a projected observable with a redshift weighting by
\begin{eqnarray}
    W_{\textrm{HI},j}\!\left(z\right) &=& b_{\textrm{HI}}(z) \bar{T}_{\textrm{HI}}(z)W_{\Delta_j}\!(z),
    \label{eqn: HI kernel}
\end{eqnarray}where the mean HI brightness temperature \citep{2006PhR...433..181F} is 
\begin{eqnarray}
\bar{T}_{\textrm{HI}}(z) = 188.8 \frac{(1+z)^2}{E(z)} \Omega_{\textrm{HI}}(z)h\ [\textrm{mK}].
\end{eqnarray}
Here, $\Omega_{\textrm{HI}}$ is the energy density fraction of neutral hydrogen, $h$ is dimensionless Hubble constant and $E(z)$ is $H(z)/H_0$.

\begin{figure}
    \centering
    \includegraphics[scale=0.37]{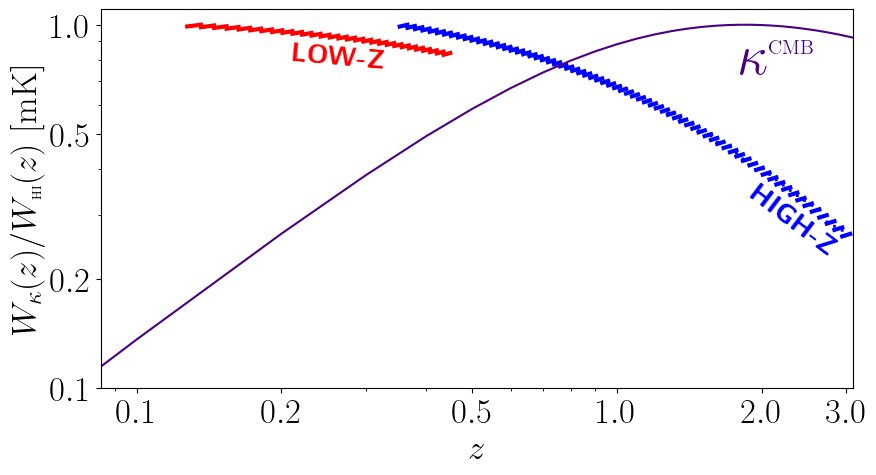}
    \caption{Window function of CMB lensing $\kappa$ described in Eq. (\ref{eqn: kappa kernel}) is shown in a purple line, and the HI window functions described in Eq. (\ref{eqn: HI kernel}) are shown as red and blue steps for the two scenarios considered in this work.}
    \label{fig: kernels}
\end{figure}

The general expression of the angular power spectrum for two generic LSS tracers $A$ and $B$, can be described as 
\begin{equation}
C^{A, B}_{\ell}(\nu_i,\nu_j) = \int \frac{\textrm{d}z H(z)}{c}\frac{W_{A,i} \left(z\right) W_{B,j} \left(z\right)}{\chi^2(z)}P\left( k, z\right),
\label{eq: LIMBER_FIELDS_A_B}
\end{equation}
which has been simplified by the Limber approximation $k=(\ell + 1/2)/\chi$ where $k$ refers to spatial and temporal coordinates and $\ell$ is the multipole. In this work, $A=\kappa$ and $B=\textrm{HI}$. We show a representative plot for the auto and cross-power spectra in Figure \ref{fig: theory}. These analyses assume full-sky coverage and a range of frequencies compatible with BINGO \citep{BINGO_I}, hereafter \textbf{LOW-Z}, and SKA1-MID \citep{ska_2020_redbook_dbacon}, hereafter \textbf{HIGH-Z}. We note that these two names refer to two different redshift ranges with an overlapping region around $z\sim 0.35$ in Figure \ref{fig: kernels}. We assume that $\Omega_{\textrm{HI}}(z)=4.86\times10^{-4}$~\citep{bull2015} and $b_{\textrm{HI}}(z)=1$. The cosmological parameter set is $(H_0, \Omega_{b}h^2, \Omega_{c}h^2, n_s) = (67.5, 0.022, 0.122, 0.965)$. The matter power spectrum is calculated via the public code CAMB\footnote{https://camb.readthedocs.io/en/latest/}. As found in~\cite{wolz2019}, the non-linear effects of the cross-power spectrum only start to dominate at very small scales up to $k\sim10 h {\rm Mpc}^{-1}$. Also, as we will show in the following text, the signal-to-noise ratios are mainly dominated by the linear scales below $\ell<200$ for the specific configurations considered in this work; thereby, the non-linear scales can be safely ignored. Because of these reasons, we only consider a linear matter power spectrum for this study. In addition, we neglected the shot-noise contributions of the power spectra as they are found to be small at all redshifts from the state-of-the-art hydrodynamic simulations~\citep{villaescusa-navarro2018,wolz2019}. In Figure \ref{fig: kernels}, the normalized redshift distributions of CMB lensing (Eq. \ref{eqn: kappa kernel}) and the HI (Eq. \ref{eqn: HI kernel}) are shown as step functions and a solid line, respectively. We note that the galaxy bias $b_{\textrm{HI}}(z, k)$ could be both redshift and scale-dependent and the HI fraction $\Omega_{\textrm{HI}}(z)$ could be redshift dependent as well. Especially, there are also non-negligible redshift-space distortions (RSDs). However, we do not include these complexities to simplify the validation of the cross-correlation algorithms. We defer the detailed investigation of how the theoretical models can affect the reconstruction of the cross-correlation to the future work.\\

\section{Simulations of the cross correlations}

\begin{figure}
    \centering
    \includegraphics[scale=0.3]{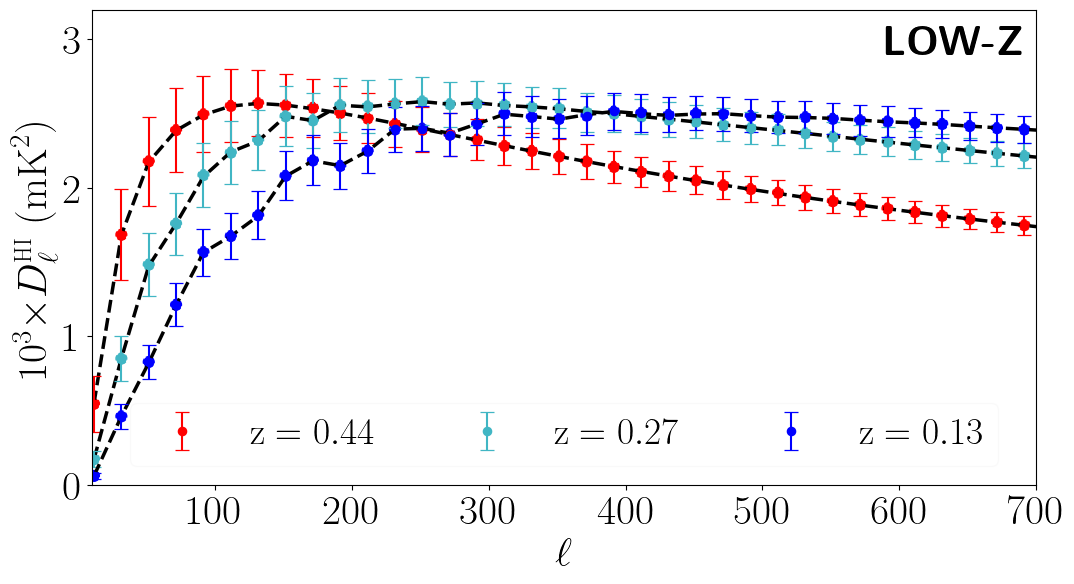} 
    \includegraphics[scale=0.3]{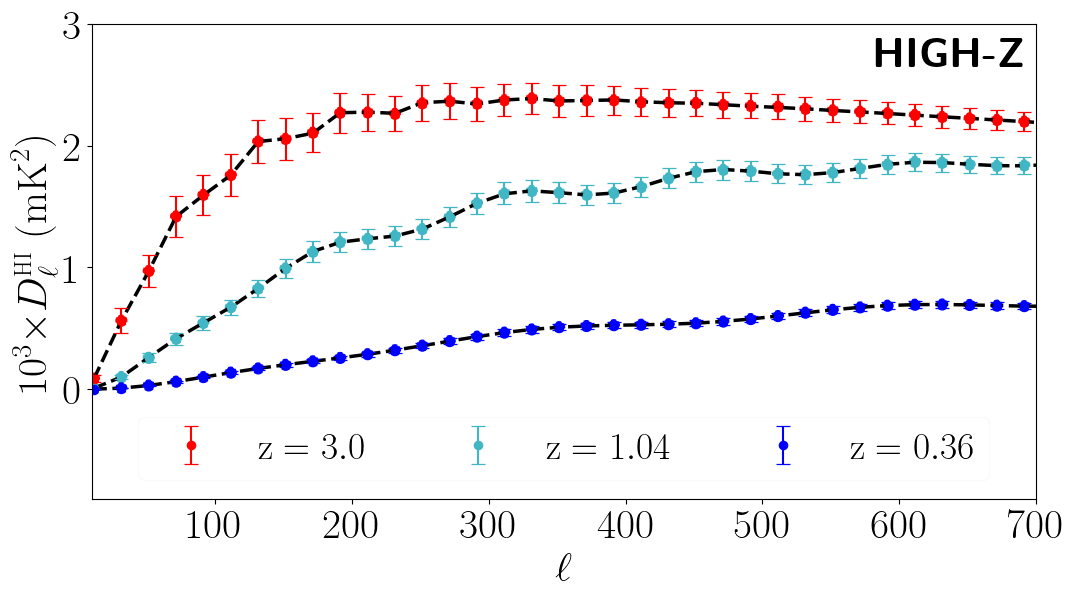}
    \caption{Validations of the correlated simulations as described in Eqs. (\ref{eq: alm_hi_sim}, \ref{eq: alm_kappa_sim}). The simulated bandpowers are consistent with the theoretical calculations, and the bandpower errors are estimated from 100 realizations. Here, we define $D_{\ell}=\ell(\ell+1)C_\ell/2\pi$.}
    \label{fig: simulations_hi_cl}
\end{figure}

To simulate correlated HI and $\kappa$ fields, we first derive all the theoretical power spectra and generate Gaussian fields with random seeds $\zeta^{j(1)}$ and $\zeta^{(2)}$ \citep{kamionkowski97}. Here, the index $j$ refers to a specific frequency. Specifically, the correlated fields are generated from   
\begin{eqnarray}
    &&a_{\ell m}^{\textrm{HI}}(\nu_j) = \sqrt{C^{\textrm{HI},}_{\ell, \textrm{fid}}(\nu_j)}\ \zeta^{j(1)}_{\ell m}\label{eq: alm_hi_sim},\\
   &&a_{\ell m}^{\kappa} = \sum_{j=1}^{n_{\textrm{M}}}\frac{C^{\textrm{HI},\kappa}_{\ell}(\nu_j)}{\sqrt{C^{\textrm{HI}}_{\ell}}(\nu_j)}\ \zeta^{j(1)}_{\ell m}\label{eq: alm_kappa_sim} + \sqrt{C^{\kappa}_{\ell} - \sum_{j=1}^{n_{\textrm{M}}}\frac{(C^{\textrm{HI},\kappa}_{\ell}(\nu_j))^2}{C^{\textrm{HI}}_{\ell, \textrm{fid}}(\nu_j)}}\ \zeta^{(2)}_{\ell m}\nonumber\\\label{GaussianSims}
\end{eqnarray}
for $n_{\rm{M}}$ HI-channels correlated with $\kappa$. The random variables satisfy the relations
\begin{eqnarray}
\bigl\langle\zeta^{j(1)}_{\ell m}(\zeta_{\ell' m'}^{k(1)})^*\bigr\rangle &=& \delta_{jk}\delta_{\ell\ell'}\delta_{m m'},\nonumber\\
\bigl\langle\zeta^{(2)}_{\ell m}(\zeta_{\ell' m'}^{(2)})^*\bigr\rangle &=& \delta_{\ell\ell'}\delta_{m m'},\nonumber\\
\bigl\langle\zeta^{j(1)}_{\ell m}(\zeta_{\ell' m'}^{(2)})^*\bigr\rangle &=& 0.
\end{eqnarray}
The lensing field from this procedure can be split into
\begin{eqnarray}
a^{\kappa}_{\ell m} &=&  \sum_{j=1}^{n_\textrm{M}}a^{ \textrm{corr}}_{\ell m}(\nu_j) + a^{\textrm{uncorr}}_{\ell m}.
\end{eqnarray}

\begin{figure}
    \centering
    \includegraphics[scale=0.3]{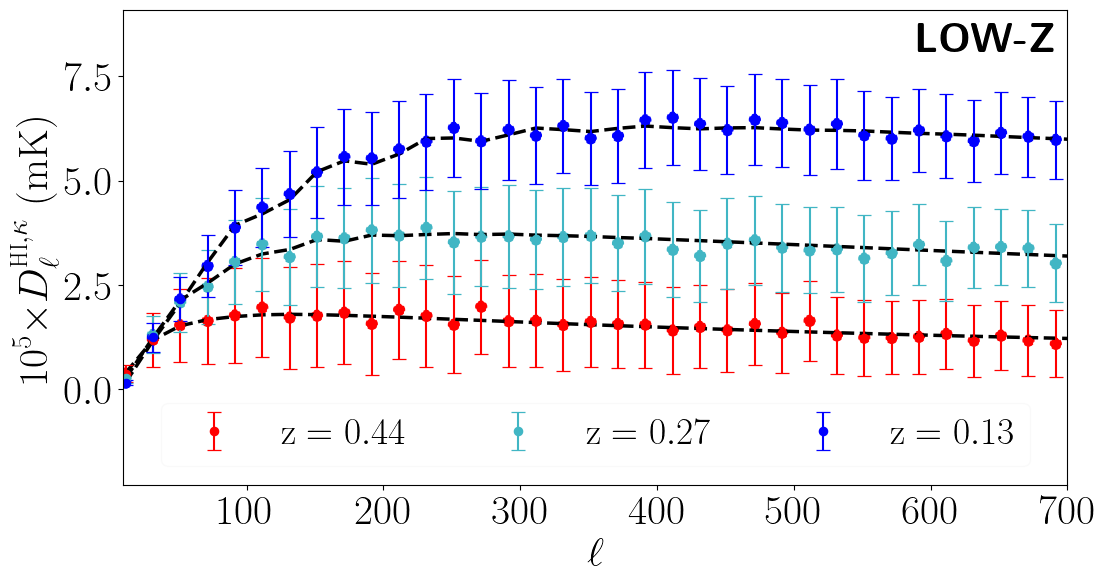}     
    \includegraphics[scale=0.3]{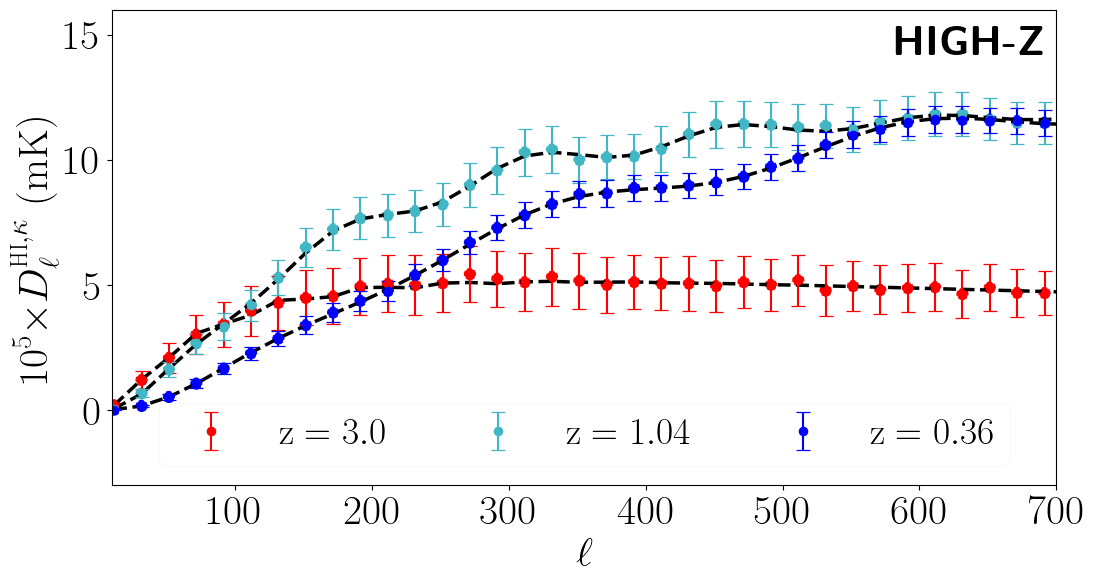}
    \caption{Validations of the cross-correlations between CMB lensing and the HI. The correlated simulations are defined in Eqs. (\ref{eq: alm_hi_sim}, \ref{eq: alm_kappa_sim}), and the error bars are estimated from 100 realizations. The error bars are multiplied by a factor of 0.2 for visualization purposes. }
    \label{fig: simulations_cross}
\end{figure}

We show the HI auto-power spectra and HI-lensing cross-power spectra that are averaged from 100 realizations in Figure \ref{fig: simulations_hi_cl} and Figure \ref{fig: simulations_cross}, respectively. These simulated power spectra agree with the input theory curves shown by the dashed black lines.\\


\section{Foreground removal}

In this work, we consider two main sources of radio foregrounds, i.e., synchrotron and free-free emissions, which are simulated by \texttt{PySM3}\footnote{https://pysm3.readthedocs.io}\citep{thorne2017}. For the free-free emission, we use a map template generated from \texttt{COMMANDER}\footnote{\href{https://github.com/Cosmoglobe/Commander}{https://github.com/Cosmoglobe/Commander}} \citep{PLANCK2016_commander}, which can extract a degree-scale map of free-free emission at 30 GHz from Planck 2015 data. The synchrotron simulation is an extrapolation from a 408 MHz synchrotron template \citep{remazeilles2015}. In addition to the Galactic foreground contamination, the extra-Galactic point sources contaminating the HI observations might also be correlated with the large-scale structure. However, as found in~\cite{2017ApJ...846...21F}, the bright radio sources can be easily identified using the existing source catalogs and removed from the maps without losing too much sky area; thus, its contamination can be mostly suppressed. Also, the extra-Galactic free-free contamination has a very smooth spectral energy distribution, and its brightness is a few orders of magnitude lower than the dominant Galactic synchrotron emission~\citep{2005ApJ...625..575S}; therefore, the blind foreground removal procedure used in this work can fully eliminate its impact on the 21 cm maps. Because of these reasons, we do not additionally consider the extra-Galactic foreground contamination.

Meanwhile, the variability of the foreground, such as transients, could also affect the component separation. However, sophisticated techniques, such as matched filters, can be applied to the time streams, and the time-domain features induced by the variable sources can be effectively removed. Alternatively, the time stream chunk with variable sources can be just flagged and removed as a conservative approach. In addition, maps from intensity mapping experiments are generated from time streams collected at a finite amount of integration time on small sky pixels that are much larger than optical surveys. The effects of the variable sources will be further averaged out.

The radio sky model at the $j$th-channel is thus
\begin{eqnarray}
x\left(\nu_j,\textbf{n}\right) 
                      &=& \sum_{k=1}^{N_{\textrm{fg}}}a_{jk}\left(\textbf{n}\right)s_k\left(\textbf{n}\right) + x_{\textrm{HI}}\left(\nu_j, \textbf{n}\right),
\end{eqnarray}where $a_{k}\left(\textbf{n}\right)$ describes the spectral energy distribution (SED) of the $k$th-foreground component, $N_{\textrm{fg}}$ is the total number of foreground components and $s_k(\textbf{n})$ is its spatial response in a given sky direction $\textbf{n}$. This model can be described with matrices
\begin{eqnarray}
\textbf{X} = \textbf{A}\textbf{S} + \textbf{X}_{\textrm{HI}},\label{radioskymat}
\end{eqnarray} where $\textbf{X} = \left[\textbf{x}\left(\nu_0\right), \textbf{x}\left(\nu_1\right),\cdots, \textbf{x}\left(\nu_{n_{\rm{ M}}-1}\right)\right]^{\textrm{T}}$ is constructed from $n_{\rm M}$ frequency channels. By minimizing the noise variance in the reconstructed map, an optimal linear combination of measurements at different frequencies can be achieved by 
\begin{eqnarray}
\widehat{\textbf{X}}_{\textrm{FG}} =\textbf{W}_{\textrm{FG}}\textbf{X},
\end{eqnarray}
with the optimal weights $\textbf{W}_{\textrm{FG}}=\widehat{\textbf{A}}\textbf{W}$ where $\textbf{W} = \left(\widehat{\textbf{A}}^{\textrm{T}}\widehat{\textbf{A}}\right)^{-1}\widehat{\textbf{A}}^{\textrm{T}}$. The radio sky model contains two major components $\textbf{X}=\textbf{X}_{\textrm{HI}}+\textbf{X}_{\textrm{FG}}$, as described by Eq. (\ref{radioskymat}). Therefore, the HI component can be obtained by subtracting the reconstructed foreground from the raw data via
\begin{eqnarray}
\widehat{\textbf{X}}_{\textrm{HI}} = \textbf{X} - \widehat{\textbf{X}}_{\textrm{FG}} = \left(\textbf{1}-\textbf{W}_{\textrm{FG}}\right)\textbf{X}.
\end{eqnarray}
With the two-component model, we can express the two reconstructed maps with the optimal weights as 
\begin{eqnarray}
 \widehat{\textbf{X}}_{\textrm{\tiny FG}} &=& \textbf{W}_{\textrm{FG}}\textbf{X}_{\textrm{FG}} + \widetilde{\textbf{L}}_{\textrm{HI}}, 
\end{eqnarray}
and
\begin{eqnarray}
 \widehat{\textbf{X}}_{\textrm{HI}} &=& \widetilde{\Delta\textbf{L}}_{\textrm{FG}} + \textbf{X}_{\textrm{HI}}-\widetilde{\textbf{L}}_{\textrm{HI}},\label{reconHI}
\end{eqnarray}
where we define the two leakage terms
$
\widetilde{\textbf{L}}_{\textrm{HI}}=\textbf{W}_{\textrm{FG}}\textbf{X}_{\textrm{HI}}$ and $
\widetilde{\Delta\textbf{L}}_{\textrm{FG}}=(\textbf{1}-\textbf{W}_{\textrm{FG}})\textbf{X}_{\textrm{FG}}$.

Eq. (\ref{reconHI}) indicates that the reconstructed HI signal is not only contaminated by foreground residuals but also a weighted sum of HI signals at all frequencies. The leakage $\widetilde{\textbf{L}}_{\textrm{HI}}$ is present due to the foreground removal procedures and is a generic contribution regardless of different blind methods. FastICA is a computationally efficient blind foreground removal algorithm, so it is adopted for this analysis. FastICA is a fast ICA fixed-point algorithm that uses negentropy to measure non-Gaussianity and forms an optimal linear combination $\widehat{\textbf{S}}=\textbf{W}\textbf{X}$, where all $n_s$-components are mutually independent \citep{hyvarinen1999}. In this work, we use the non-quadratic function $g(\cdot) = \log\cosh(\cdot)$, with 20 iterations and set one percent tolerance with FastICA provided by the scikit-learn package\footnote{\url{https://scikit-learn.org/stable/modules/generated/sklearn.decomposition.FastICA.html}}.

From Eq. (\ref{reconHI}), the estimated cross-power spectrum at each channel $j$ is
\begin{eqnarray*}
C_\ell^{\widehat{\textrm{HI}}, \kappa}(\nu_j) &=& C_\ell^{\textrm{HI}, \kappa}(\nu_j)  + \left(C_\ell^{\widetilde{\textrm{FG}}, \kappa}(\nu_j) - C_\ell^{\widetilde{\textrm{HI}}, \kappa}(\nu_j) \right)\\ &\approx& C_\ell^{\textrm{HI}, \kappa}(\nu_j)  -  C_\ell^{\widetilde{\textrm{HI}}, \kappa}(\nu_j)  \label{eq: cls_hi_rec_leak}.
\end{eqnarray*}
Here, the cross-power spectra $C_\ell^{\widetilde{\textrm{FG}}, \kappa}$ and $ C_\ell^{\widetilde{\textrm{HI}}, \kappa}$ correspond to cross correlations of $\kappa$ and the foreground leakage and that of $\kappa$ and HI leakage, respectively. The term $C_\ell^{\widetilde{\textrm{FG}}, \kappa}(\nu_j)$ vanishes since the $\kappa$ field and foreground are uncorrelated. However, the weighted sum of the HI signals due to foreground removal gives rise to an additional $\kappa$-HI correlation with a minus sign. This term for the channel $\nu$ in harmonic space is
\begin{eqnarray}
    a^{\widetilde{\textrm{HI}}}_{\ell m} (\nu) = \sum_{\nu'=1}^{n_\textrm{ch}}\omega_{\nu \nu'}a^{\textrm{HI}}_{\ell m} (\nu'),
\end{eqnarray}
where $\omega_{\nu \nu'}$ represents the components of $\textbf{W}_{\textrm{FG}}$, i.e., $\textbf{W}_{\textrm{FG}}=\{\omega_{\nu \nu'}\}$. This term implies that the leakage is due to the mixing of HI observations at different frequencies with each frequency weighted by the $\nu$-th row of $\textbf{W}_{\textrm{FG}}$.

As described in Eq. (\ref{GaussianSims}), $\kappa$ can be correlated with the HI fields at different frequencies, thus the cross-correlation of $\kappa$ and the HI leakage is not vanishing, i.e.,
\begin{eqnarray}
    C^{\widehat{\textrm{HI}},\kappa}_\ell (\nu) &\approx& C^{\textrm{HI},\kappa}_\ell (\nu) - \sum_{\nu'=1}^{n_\textrm{M}}\omega_{\nu \nu'}\big\langle a^{\textrm{HI}}_{\ell m} (\nu'),  a^{\kappa}_{\ell m}\big\rangle,
\end{eqnarray}which is just
\begin{eqnarray}
    C^{\widehat{\textrm{HI}},\kappa}_\ell (\nu) &=& C^{\textrm{HI},\kappa}_\ell (\nu) - \sum_{\nu'}\omega_{\nu \nu'} C^{\textrm{HI},\kappa}_\ell (\nu')\label{eq: cross_HIrec_expression},\\
    &=& \sum_{\nu'}\left( \delta_{\nu\nu'} - \omega_{\nu \nu'}\right) C^{\textrm{HI},\kappa}_\ell (\nu').\label{eq: cross_HIrec_expression_2}
\end{eqnarray}

This equation can be expressed in a matrix formalism as
\begin{eqnarray}
\textbf{C}_{\ell}^{\textrm{HI}\mbox{-}\kappa} =  \left(\textbf{I}- \textbf{W}_{\textrm{FG}} \right)\textbf{C}_{\ell}^{\textrm{HI}\mbox{-}\kappa, 0},
\label{eq: cross_estimated_matrix}
\end{eqnarray}
where $\textbf{C}_{\ell}^{\textrm{HI}\mbox{-}\kappa, 0}$ is the input cross-power spectrum.
The whole matrix $\textbf{W}_{\textrm{FG}}$ is sparse and may not be invertible.

\begin{figure*}[ht!]     
\includegraphics[scale=0.34]{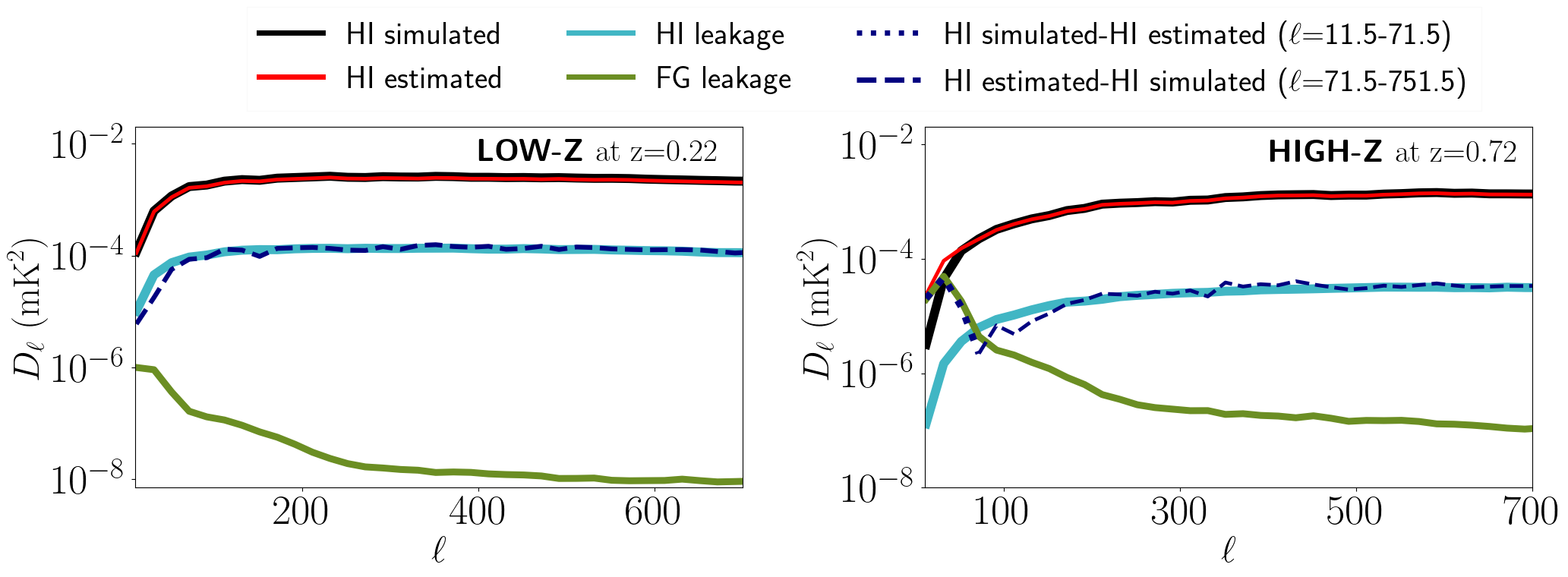}   
   \caption{Different components of the HI auto-power spectra for \lz\, (left) at $z=0.22$ and \hz\, (right) at $z=0.72$.}
   \label{fig: Dl_hi_leakage}
\end{figure*}

\begin{figure*}
   \centering
   \includegraphics[scale=0.41]{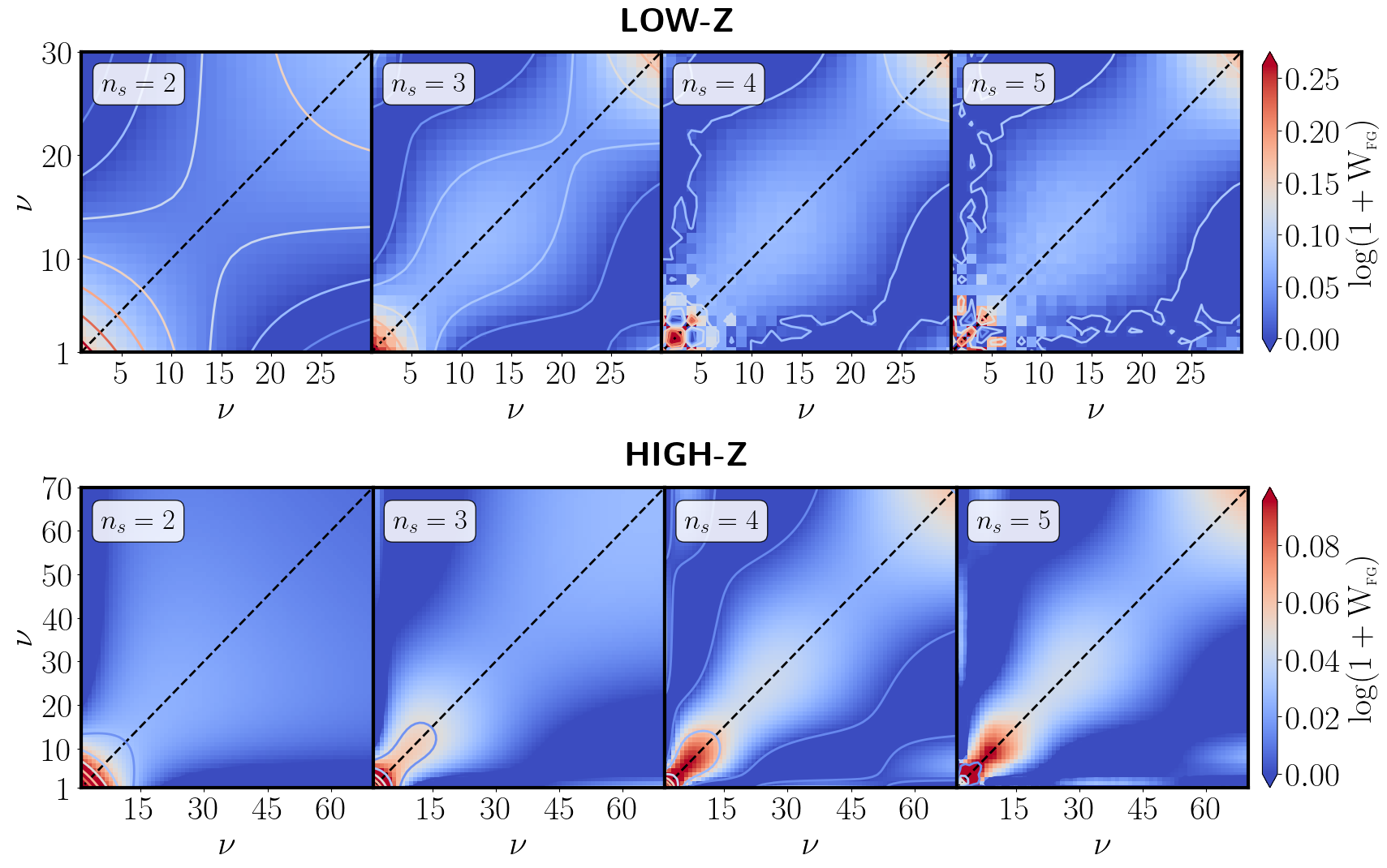}   
   \caption{The mixing matrix $\textbf{W}_{\textrm{\tiny FG}}$ due to the foreground removal is obtained via the FastICA method. The subpanels are the matrices with different numbers of independent components ($n_s$). The x and y axes correspond to the matrix dimension related to the HI channel indices, and the color bar reflects the relative values of the mixing matrices.}
   \label{fig: WFG_filter}
\end{figure*}

As presented in \cite{cunnington2019}, an approximation of a constant $\textbf{W}_{\textrm{FG}}$ mixing matrix for a foreground removal procedure was adopted to avoid spurious effects from blind algorithms, which cannot distinguish the radial models of the foreground from that of the HI. The authors showed that removing the mean HI fluctuations along the line-of-sight cannot effectively mitigate the edge effect caused by blind foreground removals but can conserve the spectrum shape and partially recover its power spectrum.

Removing the mean HI fluctuations along the line of sight is a good approximation of the procedure described in Eq. (\ref{reconHI}), especially when we are interested in cross-correlation analysis, as shown in Figure \ref{fig: Dl_hi_leakage}. This approach assumes that HI leakage is an average among all frequency channels. Therefore, the reconstructed HI map in Eq. (\ref{reconHI}) becomes
\begin{eqnarray}
    \widehat{\textbf{X}}_{\textrm{HI}} = \textbf{X}_{\textrm{HI}}  - \widetilde{\textbf{L}}_{\textrm{HI}} \approx \textbf{X}_{\textrm{HI}}  - \bar{\textbf{X}}_{\textrm{HI}}.
    \label{eq: cunnington_approach}
\end{eqnarray}
The second term refers to the average over all frequencies. In particular, the HI brightness temperature in \cite{cunnington2019} is
\begin{eqnarray}
    \delta\widehat{T}_{\textrm{HI}}(\textbf{n},\nu)  &=& \delta T_{\textrm{HI}}(\textbf{n},\nu)   - \delta\bar{T}_{\textrm{HI}} (\textbf{n})\nonumber\\
    &=& \delta T_{\textrm{HI}}(\textbf{n},\nu)  - \frac{1}{n_{\textrm{ch}}}\sum_{\nu'}\delta T_{\textrm{HI}}(\textbf{n},\nu'),
        \label{eq: HI_estimation_anut_approach}
\end{eqnarray}
which is similar to our procedure in Eq. (\ref{reconHI}), except that they chose a uniform weighting $\omega_{\nu\nu'}=1/n_{\textrm{ch}}$ for all $\nu'$, whereas we derive this foreground removal efficiency matrix precisely from simulations.

We calculate the mixing matrix from a noiseless radio sky model with foreground components that consist of free-free and synchrotron emission. The results are shown in Figure \ref{fig: WFG_filter}, where the $\textbf{W}_{\textrm{FG}}$ matrices are computed for $n_{\textrm{s}}=2-5$ independent components for the \textbf{LOW-Z} and \textbf{HIGH-Z} scenarios. We find that the first and last channels in the mixing matrix have spuriously higher values than the remaining channels because of limited information at the boundaries, which is exactly the edge effect.\\
 
\begin{figure*}
   \centering
   \includegraphics[scale=0.33]{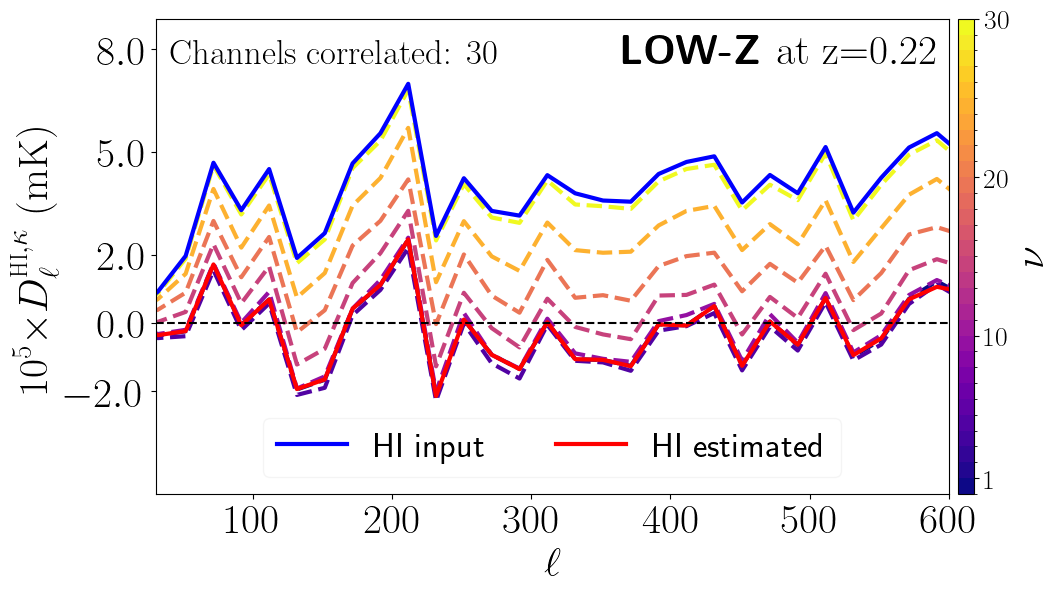}
   \includegraphics[scale=0.33]{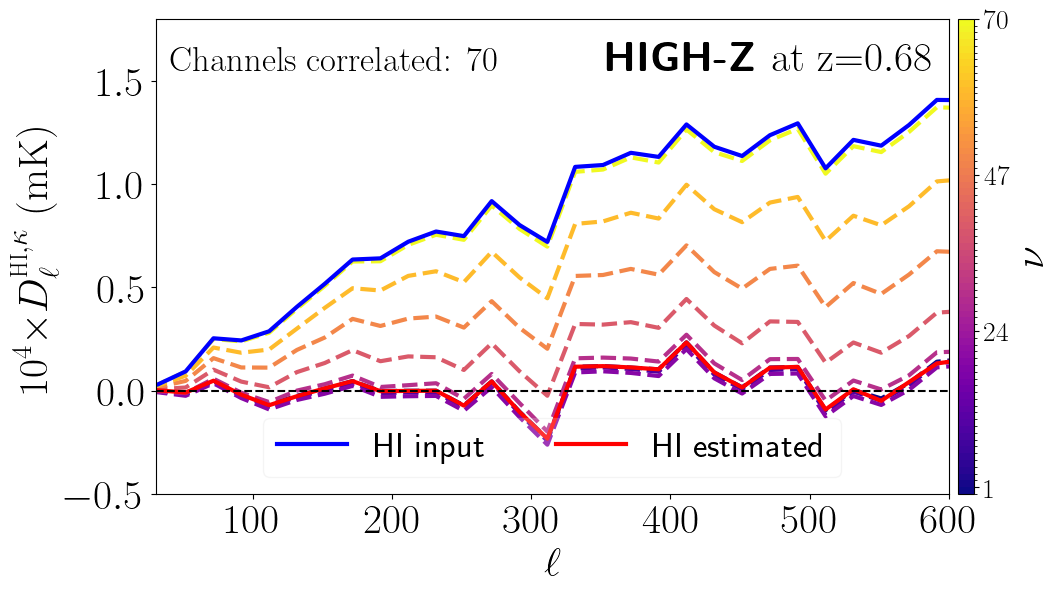}   
   \caption{Cross-correlations between HI and $\kappa$ with the HI simulated directly from input theory (blue solid) and the reconstructed HI from FastICA (red solid). We assume that $\kappa$ is correlated with all HI channels. The left plot shows the cross-power spectrum of a specific frequency channel, i.e., the 20th channel, for the \lz\, scenario. The right plot shows the cross-power spectrum of the 48th channel for the \hz\, scenario. The colored dashed lines are calculated from Eq. (\ref{eq: cross_HIrec_expression}) with varying the effective number of frequency channels. The results are obtained with $n_s=4$.}
   \label{fig: Dl_cx_all_channels}
\end{figure*}

\section{Impact of foreground removal on the cross correlations}

Eq. (\ref{eq: cross_HIrec_expression_2}) indicates that each channel contributes to the cross-power spectrum. The effects of the leakage term are shown in Figure \ref{fig: Dl_cx_all_channels} for \textbf{LOW-Z} and \textbf{HIGH-Z}. The dashed curves from blue to yellow are the contributions from different numbers of mixed channels, i.e., $\sum_{\nu'=1}^{N_{\rm eff}}\omega_{\nu \nu'}C^{\textrm{HI},\kappa}_\ell (\nu')$ where $N_{\rm eff}$ is gradually increased from $N_{\rm eff}=2$ to $N_{\rm eff}=n_M$. When $N_{\rm eff}=n_M$, the leakage term is almost the same as the solid blue line which is the input theory curve. This is because the blind approaches only minimize the variances of the foreground components, resulting in suboptimal reconstructions of the HI signal. There might be new schemes to simultaneously optimize both the foreground and HI reconstructions, achieving minimal cross-correlation leakages. However, this discussion is beyond the scope of this work.

\begin{figure*}
   \centering
   \includegraphics[scale=0.35]{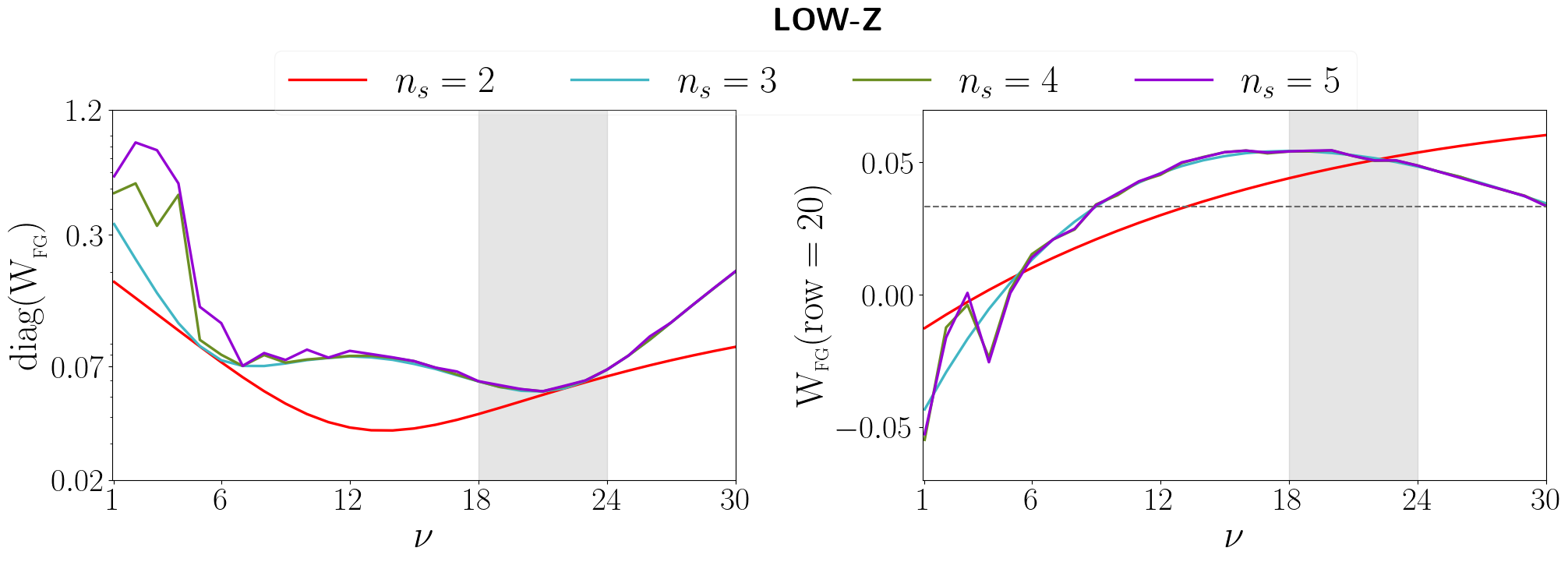}
   \caption{Mixing matrix elements of $\textbf{W}_{\textrm{FG}}$ for the \lz\, scenario. The left plot represents the diagonal matrix elements for 2-5 independent components. The right plot shows the matrix elements from the 20th row of the mixing matrix described in the second term of Eq. (\ref{eq: cross_HIrec_expression}). The gray-shaded regions represent a subset of frequencies tested for the leakage suppression. The dashed black line in the right panel represents the uniform weighting with an amplitude of 1/30.}
   \label{fig: Wfg_ns_low_redshifts}
\end{figure*}

\begin{figure*}
   \centering
   \includegraphics[scale=0.35]{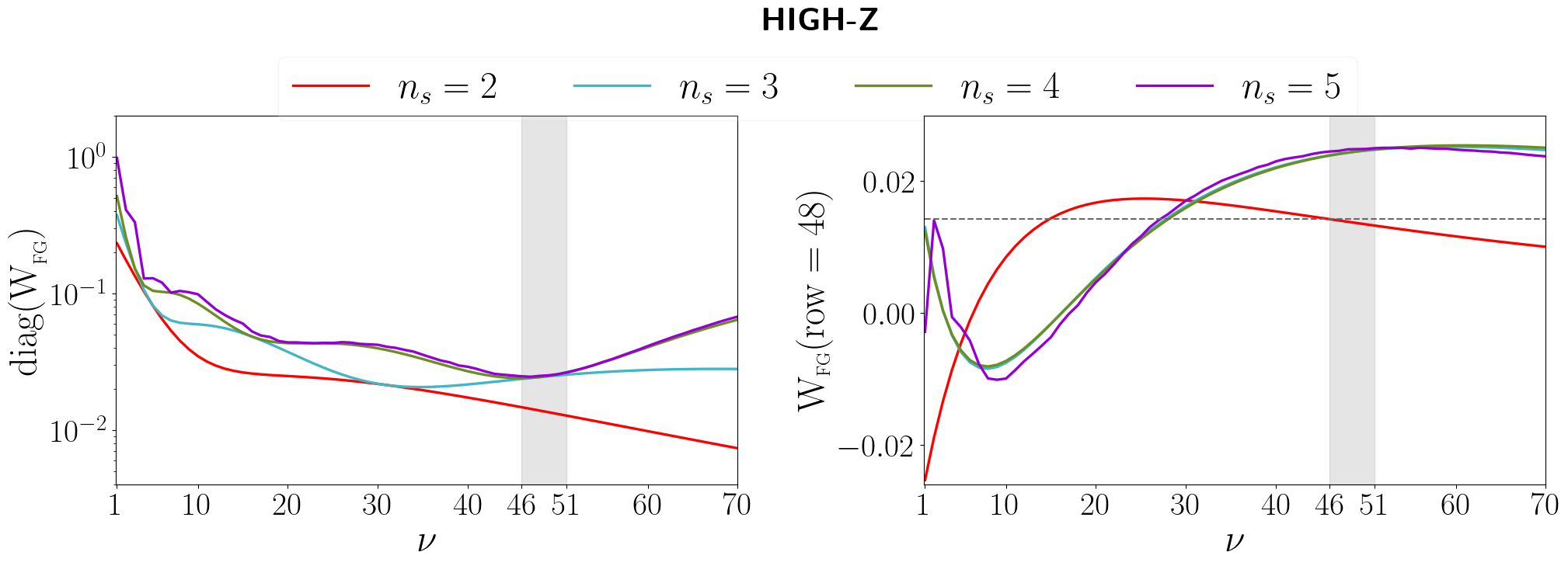}
   \caption{Mixing matrix elements of $\textbf{W}_{\textrm{FG}}$ for the \hz\, scenario. The left plot represents the diagonal matrix elements for 2-5 independent components. The right plot shows the matrix elements from the 48th row of the mixing matrix described in the second term of Eq. (\ref{eq: cross_HIrec_expression}). The gray-shaded regions represent a subset of frequencies tested for the leakage suppression. The dashed black line in the right panel represents the uniform weighting with an amplitude of 1/70.} 
   \label{fig: Wfg_ns_high_redshifts}
\end{figure*}

In light of the relationship between leakage and the number of frequency channels, one straightforward way to reduce leakage is to reduce the effective number of frequencies, for example, by correlating only the lensing map with a subset of HI frequency channels. In Figure \ref{fig: Wfg_ns_low_redshifts} and Figure \ref{fig: Wfg_ns_high_redshifts}, we show the diagonal elements and a particular row of the mixing matrix (Figure \ref{fig: WFG_filter}) in the left and right plots, respectively. The mixing matrices are calculated by varying the independent ICA modes from $n_s = 2$ to $n_s=5$. In comparison, the right plot indicates that the weighting of each frequency after component separation is different from a uniform weighting, which would cause biases in the cross-power spectrum estimation. 

For \textbf{LOW-Z}, neighboring frequencies around channel 20 can reach the same level of mixing with different numbers of independent components, as shown in Figure \ref{fig: Wfg_ns_low_redshifts}. Therefore, we choose a subset of the frequency channels between 18 and 24 labeled by the gray region. For \textbf{HIGH-Z}, the edge effect is clearly observed from the first few channels corresponding to higher redshifts in Figure \ref{fig: Wfg_ns_high_redshifts}. Similar to the discussions for the \textbf{LOW-Z}, the region between channels 46 and 51 can reach roughly the same level of mixing, so we choose the median frequency as the 48th channel. The exact mixing obviously deviates from a uniform mixing that is inversely proportional to the number of frequencies, i.e., $\omega_{\nu\nu'}\sim n_M^{-1} \approx 0.01$ when $n_M$=70. 

\begin{figure*}
   \includegraphics[scale=0.34]{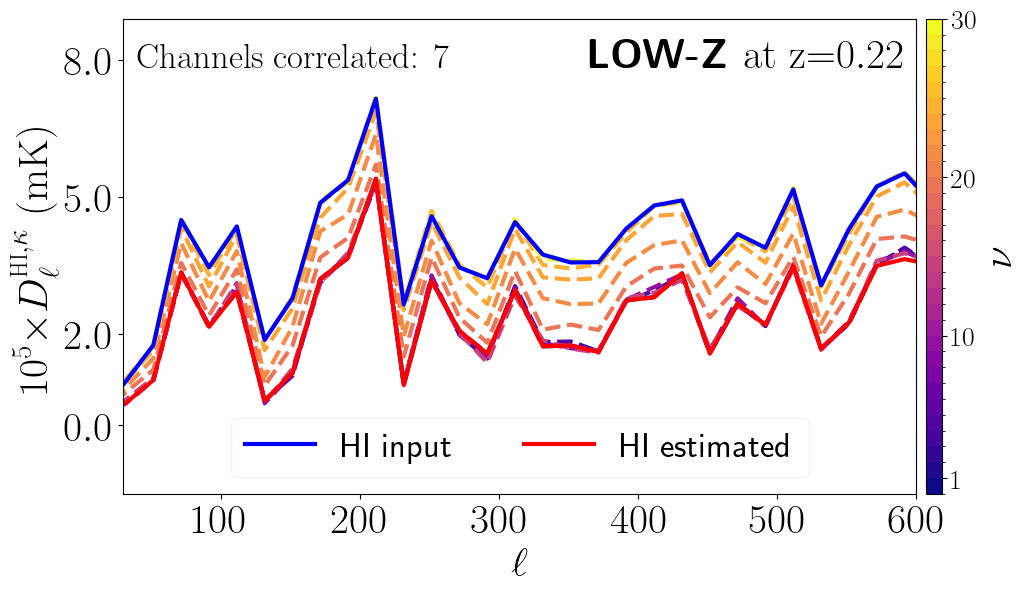}
   \includegraphics[scale=0.34]{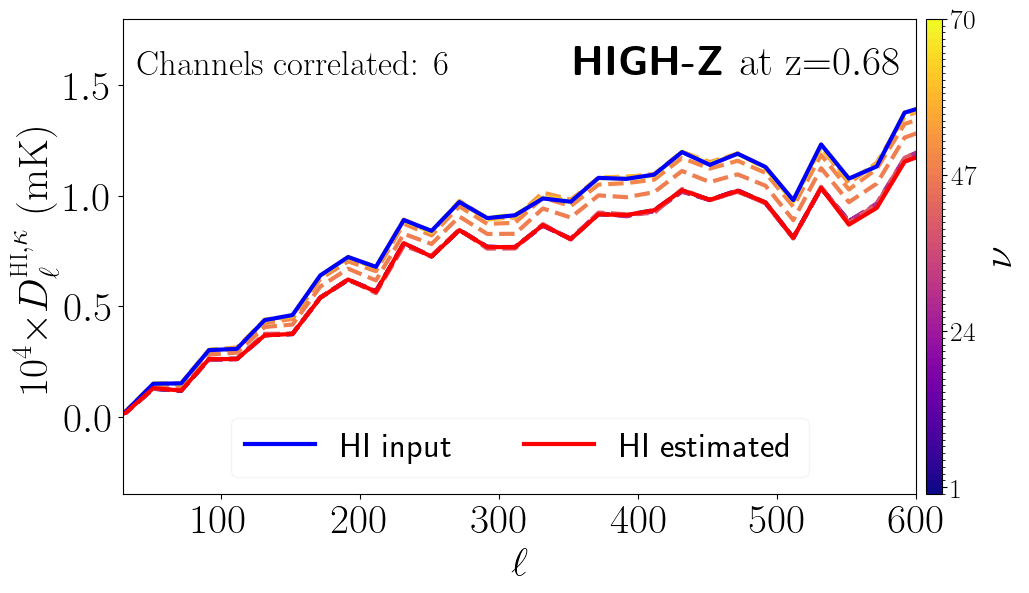}     
   \caption{Cross-correlations between HI and $\kappa$ with the HI simulated directly from input theory (blue solid) and the component-separated HI (red solid). We assume that the $\kappa$ field is only correlated with HI channels at a subset of frequencies. The left plot shows the cross-power spectrum of the 20th channel for the \lz\, scenario, and the right plot shows the cross-power spectrum of the 48th channel for the \hz\, scenario. The colored dashed lines are different leakage terms calculated from Eq. (\ref{eq: cross_HIrec_expression}) with varying the effective number of frequency channels. The results are obtained with $n_s=4$.}
   \label{fig: Dl_cross_ncorrelated_channels_part2}
\end{figure*}

Therefore, if we assume that the lensing field is only correlated with a subset of the HI channels, the resulting cross-power spectra will become nonvanishing since the leakage terms for such a scenario are greatly suppressed, as shown in Figure \ref{fig: Dl_cross_ncorrelated_channels_part2}. This scenario can be easily achieved if the CMB lensing field is replaced by a cosmic shear field or a narrow tomographic bin of the LSS tracer.

However, if the leakage term $\sum_{\nu'=1}^{N_{\rm eff}}\omega_{\nu \nu'}C^{\textrm{HI},\kappa}_\ell (\nu')$ can be precisely estimated, the cross-power spectrum between the CMB lensing and HI can still be measured after debiasing the leakage contributions even if all HI frequency channels are correlated with the CMB lensing.

We use both numerical simulations and theoretical calculations to estimate the leakage of the cross-correlation. Since the mixing matrix only depends on the foregrounds, we can assume that it is not dependent on HI and $\kappa$ realizations, i.e., $\langle\omega_{\nu\nu'}\rangle\approx\omega_{\nu\nu'}$. We calculated the mixing matrices from both a single realization and 100 realizations and found that the difference is negligible. The leakage term for a specific frequency channel $j$ is estimated by an average over realizations
\begin{eqnarray}
\langle \hat L_{\ell}^{ \textrm{\tiny HI}(\nu_j) ,\kappa}\rangle_{\textrm{sim}} = \sum_{\nu'}\omega_{\nu_j\nu'}\langle C_{\ell}^{ \textrm{\tiny HI} ,\kappa}(\nu')\rangle.
\label{eq: leakage_cx_sim}
\end{eqnarray}

In the limit of a sufficiently large number of simulations, the expression above converges to the theoretical expectation 
\begin{eqnarray}
\langle \hat L_{\ell}^{ \textrm{\tiny HI}(\nu_j) ,\kappa}\rangle_{\textrm{th}} = \sum_{\nu'}\omega_{\nu_j\nu'}C_{\ell,\textrm{fid}}^{ \textrm{\tiny HI} ,\kappa}(\nu').
\label{eq: leakage_cx_theory}
\end{eqnarray}
Here, \emph{fid} refers to the fiducial model. 
\begin{figure*}
   \includegraphics[scale=0.33]{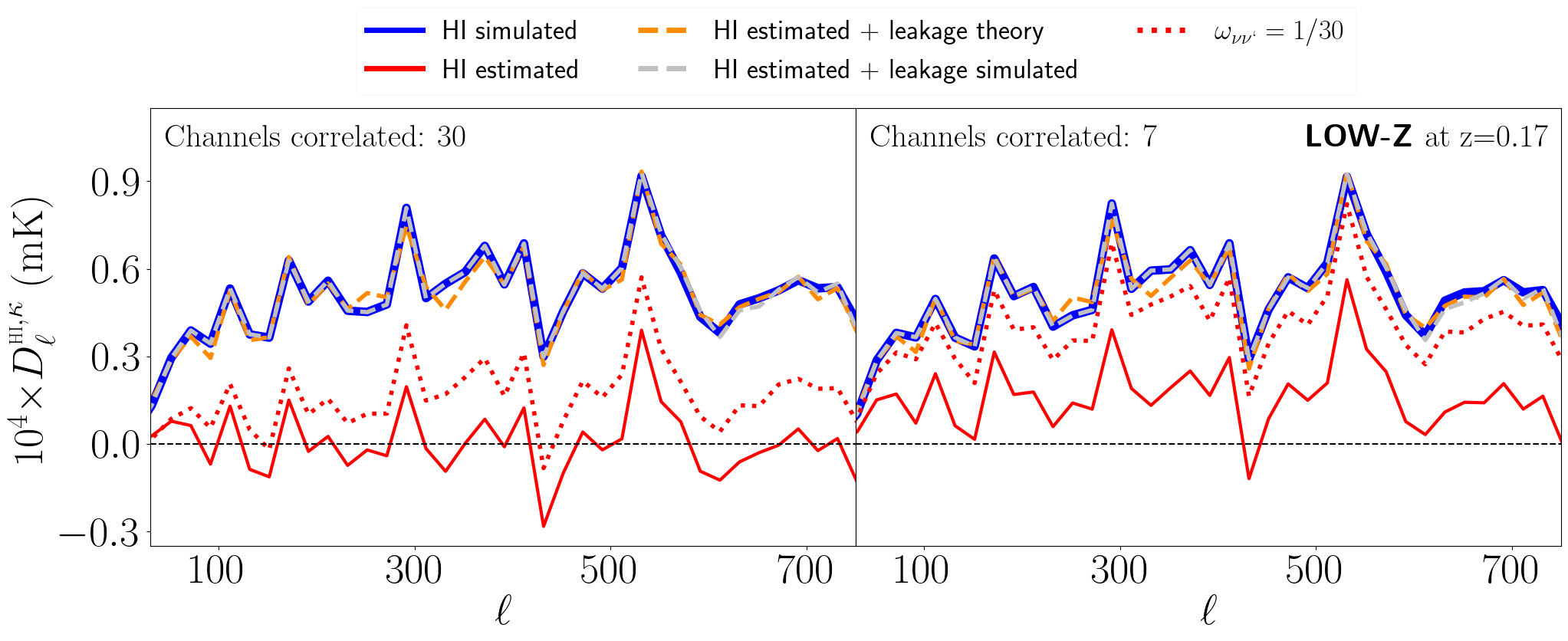}
   \caption{Angular cross-power spectra for the \textbf{LOW-Z} configuration with an effective redshift of 0.17 (channel 26) when $\kappa$ is correlated with both all the HI-channels (left) and only seven channels (right), i.e., channels 18-24. In addition to the cross-power spectra with HI before (blue) and after (red) the foreground removal procedure, we also show the debiased cross-power spectra with Eq. (\ref{eq: leakage_cx_sim}) (gray) and Eq. (\ref{eq: leakage_cx_theory}) (orange), as well as a uniform weighting scheme (dashed red). The results are obtained with $n_s=4$.}
   \label{fig: leakage_correction_lowz}
\end{figure*}

\begin{figure*}
   \includegraphics[scale=0.33]{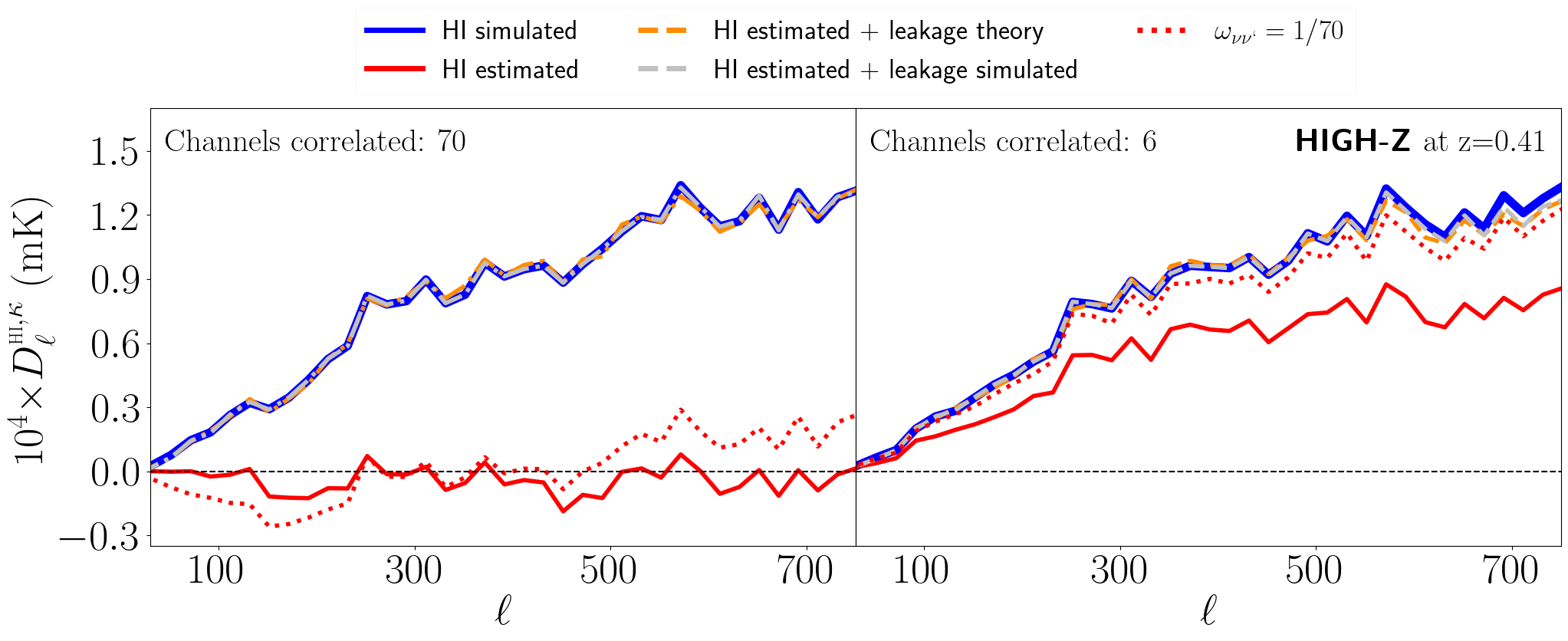}
   \caption{Angular cross-power spectra for the \textbf{HIGH-Z} configuration at an effective redshift of 0.41 (channel 66) when $\kappa$ is correlated with both all the HI-channels (left) and only six channels (right), i.e., channels 46-51. In addition to the cross-power spectra with HI before (blue) and after (red) the foreground removal procedure, we also show the debiased cross-power spectra with Eq. (\ref{eq: leakage_cx_sim}) (gray) and Eq. (\ref{eq: leakage_cx_theory}) (orange), as well as a uniform weighting scheme (dashed red). The results are obtained with $n_s=4$.}
   \label{fig: leakage_correction_highz}
\end{figure*}

To mimic an LSS tracer from a galaxy survey, we show the leakage estimations for the 26th channel ($z\sim0.17$) of \lz\, in Figure \ref{fig: leakage_correction_lowz} and the 66th channel ($z\sim0.41$) of \hz\, in Figure \ref{fig: leakage_correction_highz}. The solid blue curves, which are the $\kappa$-HI cross-power spectra from the input simulations, are consistent with the theoretical predictions. The left plot shows the tests with all HI channels correlated with $\kappa$, and the right plot is the case when only a few HI channels are correlated with $\kappa$. Obviously, the cross-power spectrum for the latter case becomes non-vanishing. \\

\section{Experimental configurations}
In the previous section, we investigated the cross-correlation leakages for cases with both all channels and a subset of HI observations correlated with $\kappa$. To assess the detectability of the cross-power spectra, we assume a \lz\, experiment similar to the BINGO telescope \citep{BINGO_III} and a \hz\, experiment similar to the SKA-MID band 1 \citep{ska_2020_redbook_dbacon}. We assume a lensing template from next-generation CMB experiments such as CMB-S4 \citep{CMBS4_2019} and a cosmic shear template from the Rubin Observatory's Legacy Survey of Space and Time (LSST) \citep{LSST2018}. 

The HI IM noise power spectrum is assumed to be $N_{\ell}=\sigma^2_{\rm pix}\Omega_{\rm pix}$, which is further deconvolved by a Gaussian beam  
$
b_{\ell}=\exp[-\ell(\ell+1)\theta^2_{ \textrm{{FWHM}}}/(16\log{2})]$ with a resolution $\theta_{ \textrm{FWHM}}$. The sensitivity of a pixel is determined by
\begin{eqnarray}
    \sigma_{\textrm{pix}} = \textrm{K}\frac{T_{\textrm{sys}}}{\sqrt{\Delta\nu t_{\textrm{pix}}}} ,
    \label{eq: HI white noise}
\end{eqnarray}
where K is noise performance, $T_{\textrm{sys}}$ is a system temperature, and $\Delta\nu$ is a bandwidth. The pixel integration time is 
\begin{eqnarray}
    t_{\textrm{pix}}= \epsilon t_{\textrm{sur}}N_{\textrm{beams}}\frac{\Omega_{\textrm{pix}}}{\Omega_{\textrm{sur}}}.
\end{eqnarray}
Here, $\Omega_{\textrm{sur}}$ is the survey coverage area, $\epsilon$ is the coverage efficiency, $N_{\textrm{beams}}$ is the number of beams, and the pixel area is assumed to be $\Omega_{\textrm{pix}}\approx\pi\theta^2_{\textrm{\tiny FWHM}}/4$. 
\begin{table}
\footnotesize
\centering
\caption{Experimental specifications for the HI IM surveys investigated in this work. Two typical scenarios are considered.}
\begin{tabular}{l c c }
\cline{1-3}
  & \lz & \hz \\
 \hline
Angular resolution ($\theta_{ \textrm{\tiny FWHM}}$)  & 40' & 2$^\circ$\\ 
Frequency range      & $980-1260$ MHz  & $350-1050$ MHz \\
Number of channels ($\textrm{n}_\nu$)  & $30$            & $70$  \\
Bandwidth ($\Delta\nu$)  & $9.33$ MHz  & $10$ MHz  \\
Sky coverage ($\Omega_\textrm{sur}$)   & 41253 deg$^2$ & 41253 deg$^2$ \\
Number of beams ($N_{\textrm{beams}}$) & 1084 & 270  \\
System temperature ($\textrm{T}_\textrm{sys}$)  & $70$ K          & $24-62$ K   \\
Total coverage time ($t_\textrm{sur}$)  & 1 year  & 1 year \\
Coverage efficiency ($\epsilon$) & 1 & 1 \\
Noise performance (K) & $\sqrt{2}$ & 1 \\
\hline
Noise per pixel ($\sigma_{\textrm{pix}}$)  &  60 $\mu$K  & 80-204 $\mu$K\\
\cline{1-3}
\end{tabular}
\label{tab: requirement}
\end{table}
White noise realizations are drawn from the noise power spectra calculated with the parameters listed in Table \ref{tab: requirement}.  

As found in \cite{BINGO_III} and \cite{Matshawule}, sophisticated optical designs can achieve approximately Gaussian beams with low sidelobes for intensity mapping experiments such as BINGO and SKA-MID/MeerKAT. Therefore, in this work, we assume a frequency-independent Gaussian beam profile with a constant resolution. Frequency-dependent beam uncertainties can undermine smooth frequency correlations of bright foreground contaminants, thereby degrading the performance of foreground removal procedures. However, for intensity mapping experiments targeting the post-reionization Universe, the frequency ranges are usually narrow and the ratios of two different frequencies are close to unity, thus the beam resolution differences among different frequency channels can be approximated as beam calibration errors. Details of real optical designs and constructions of the telescopes are required to accurately model beam shapes, and this investigation is beyond the aims of this work. Therefore, we defer the detailed investigations of systematic issues, including the 1/$f$ noise and beam chromaticity, to the future work.

The CMB lensing can be reconstructed with quadratic estimators \citep{waynehu2007}. In this work, we consider only the $EB$ estimator since its Gaussian bias is the smallest for high-sensitivity polarization experiments \citep{waynehu2002}. Different Gaussian biases can be calculated via \texttt{quicklens} code\footnote{\href{https://github.com/dhanson/quicklens/}{https://github.com/dhanson/quicklens/}}. We assume a CMB-S4-like noise level, i.e., $\sqrt{2}$ $\mu$K arcmin in polarization, and a $1'$ Gaussian beam \citep{CMBS4_2019} for a full sky coverage. 

For a galaxy-lensing-like survey, the shot noise contribution should be considered due to the finite sample of resolved sources and can be expressed as   
\begin{eqnarray} 
\sigma_{\textrm{N}}=\sigma_{\epsilon}\sqrt{\frac{A_{\textrm{pix}}N_{\textrm{bins}}}{n_{\textrm{gal}}}},
\label{eq: sigma_pix_gal}
\end{eqnarray}
where $\sigma_{\epsilon}$ is the standard deviation of the observed ellipticities in the survey, $n_{\textrm{gal}}$ is the surface number density of galaxies (per angular resolution), $A_{\textrm{pix}}$ is the angular resolution, and $N_{\textrm{bins}}$ is the number of bins. We use the LSST parameter set \citep{LSST2018} but consider a full sky survey with a set of experimental parameters $(\sigma_{\epsilon}^2, n_{\textrm{gal}}/ [\textrm{gal/arcmin}^2], A_{\textrm{pix}}/[\textrm{arcmin}^2], N_{\textrm{bins}}) = (0.26, 27, 5.49, 6)$. 

The covariance matrix for the cross-power spectrum can be expressed as~\citep{Lloyd1995} 
\begin{eqnarray}
\Delta^2 (\hat{C}_{\ell}^{A,B})&=& \frac{1}{(2\ell +1)f_{\textrm{sky}}\Delta\ell} \Big\{  \hat{C}_{\ell}^{A,B}\hat{C}_{\ell}^{A,B}+\nonumber\\ 
&& \Big[\hat{C}_{\ell}^{A}+\frac{\mathcal{N}_{\ell}^{A}}{(b_{\ell}^{A})^2}\Big]\Big[\hat{C}_{\ell}^{B}+\frac{\mathcal{N}_{\ell}^{B}}{(b_{\ell}^{B})^2)}\Big] \Big\},
\label{eq: fields_AB_variance}
\end{eqnarray}
where $b_{\ell}^{\mathcal{X}}$ is the beam profile for an observable $X$, $\mathcal{N}_{\ell}^{\textrm{}{X}}$ is its noise power spectrum, $f_{\rm sky}=1$ and $\Delta \ell=20$. Thus, the signal-to-noise ratio (SNR) per multipole is
\begin{eqnarray}
   \textrm{SNR}_{\textrm{A},\textrm{B}}(\nu,\ell) = \sqrt{\Bigg
\langle\frac{\hat{C}_{\ell}^{A,B}(\nu)}{\Delta \hat{C}_{\ell}^{A,B}(\nu)}\Bigg\rangle^2},
    \label{eq: SN_multipole}
\end{eqnarray}
with $\Delta \hat{C}_{\ell}^{A,B}$ given by Eq. (\ref{eq: fields_AB_variance}), and the brackets indicate an average over the realizations. In this work, $A=\kappa$ and $B=$ HI. We have assumed FastICA with $n_s=4$ for all the cases since they provided equivalent results from  $n_s=3-5$ for \lz\, and the optimal result for \hz.

\begin{figure*}[ht!]
   \centering
   \includegraphics[scale=0.29]{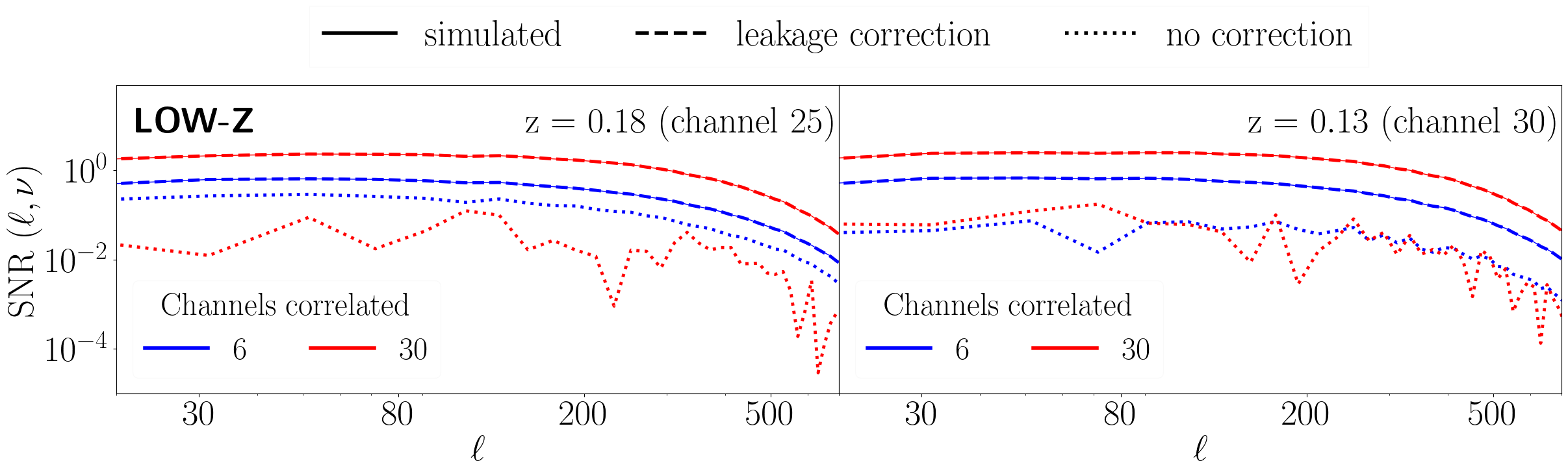}  
   \\   
    \includegraphics[scale=0.29]{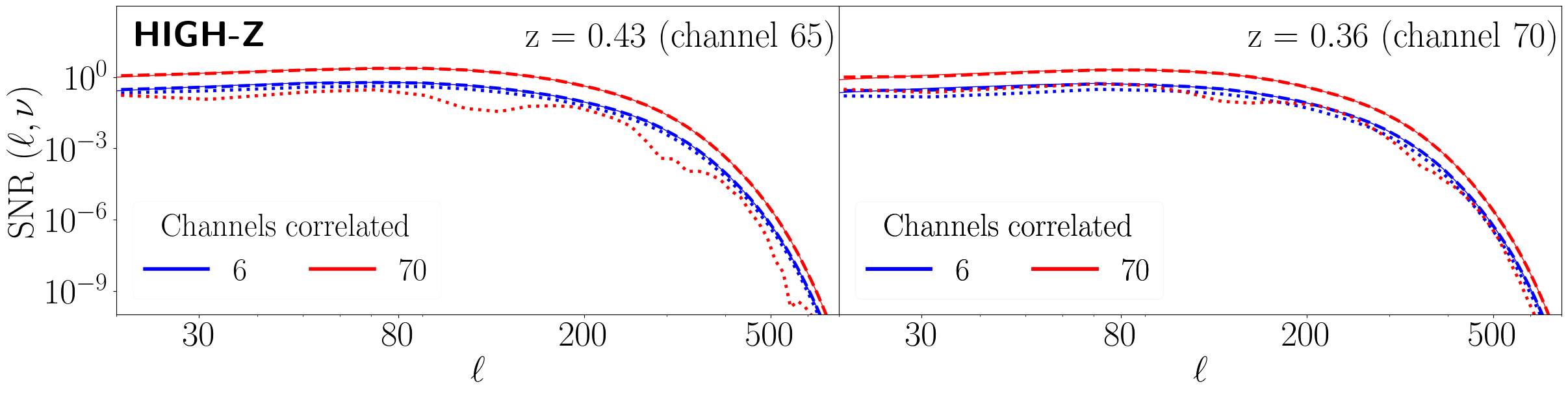}      
   \caption{The averaged signal-noise ratios (SNRs) per multipole (Eq. \ref{eq: SN_multipole}) from 100 realizations. In this figure, we show the SNRs at two different redshifts for both the \lz\, and \hz\, scenarios. Especially, we compare the SNRs for cross-correlations with the CMB lensing in red and with tomographic tracers in blue. }
   \label{fig: SN_multipoles}   
\end{figure*}

The SNRs for different scenarios are shown in Figure \ref{fig: SN_multipoles}. The main difference between \lz\ and \hz\ is caused by the resolutions of the HI surveys, and the SNRs of \lz\ decrease more slowly than those of \hz. The SNRs can be further improved once the leakage terms are estimated and the corrections are applied. 

\begin{figure*}
   \centering
   \includegraphics[scale=0.37]{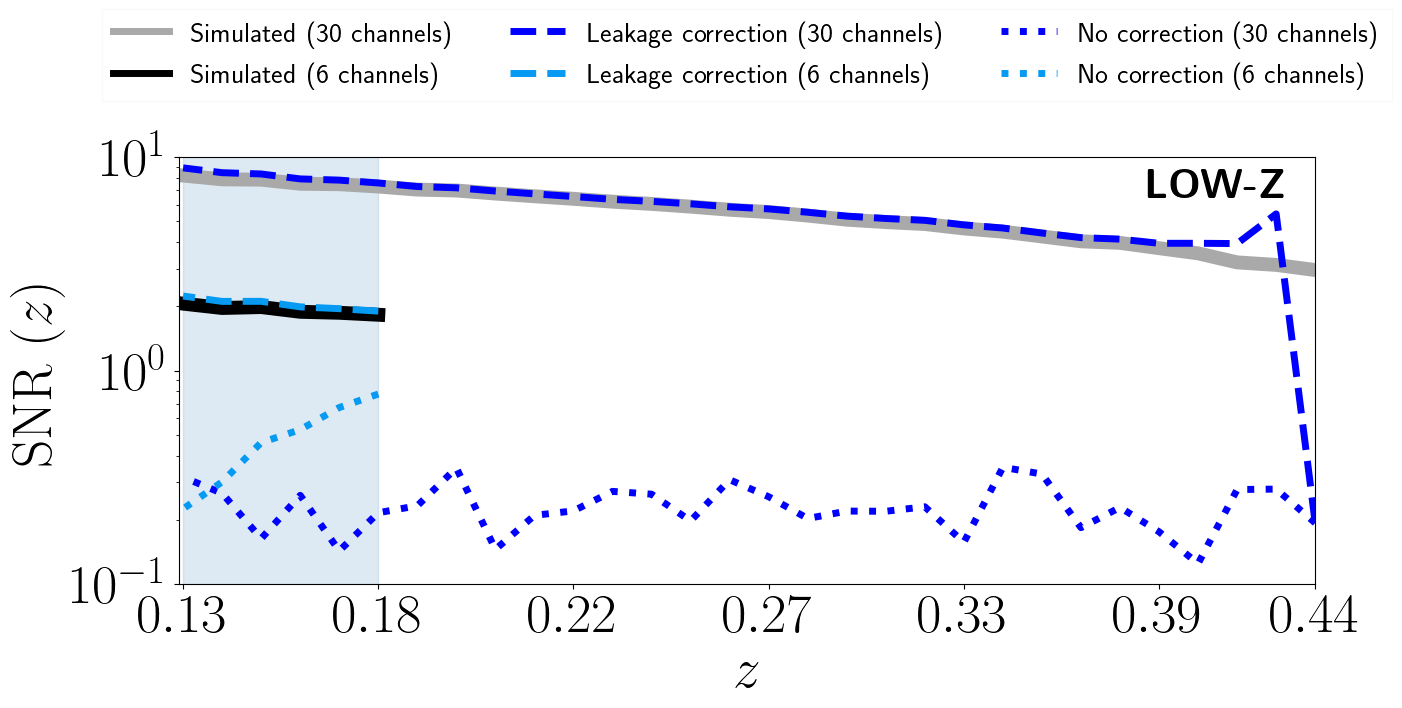}\\
   \includegraphics[scale=0.37]{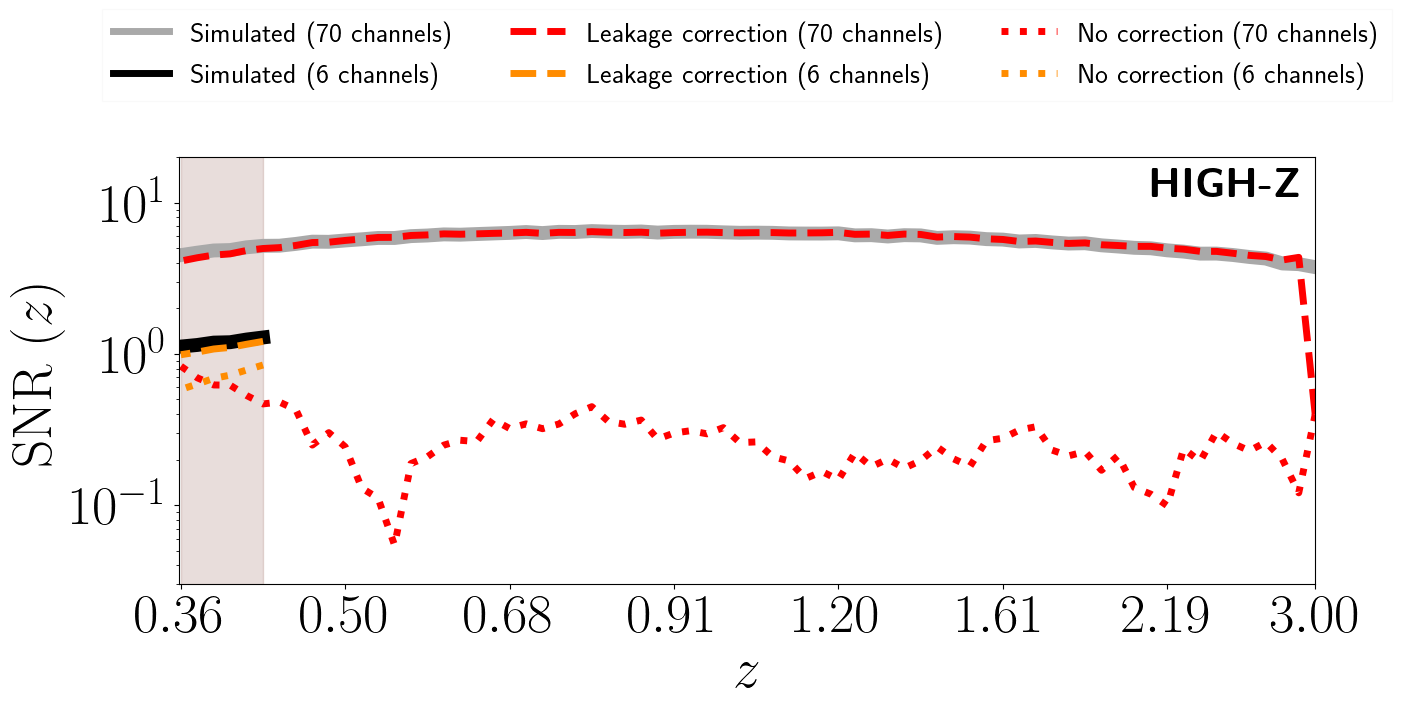}       
   \caption{The signal-to-noise ratios per effective redshift or frequency channel. The effective redshift is taken as the mean redshift of a given channel. The shaded regions represent the galaxy-survey-like scenario where only a few HI frequency channels are correlated with the CMB lensing.}
   \label{finalSNRs}   
\end{figure*}

In addition to the SNRs per multipole, we can also obtain the cumulative SNRs for each frequency band (or a tomographic band), i.e., $\textrm{SNR}_{\textrm{A},\textrm{B}}(\nu) = \sqrt{\sum_\ell \textrm{SNR}^2_{\textrm{A},\textrm{B}}(\nu,\ell)}$ as shown in Figure \ref{finalSNRs}, and even an overall SNR as $\textrm{SNR}_{\textrm{A},\textrm{B}}= \sqrt{\sum_{\nu,\ell} \textrm{SNR}^2_{\textrm{A},\textrm{B}}(\nu,\ell)}$ listed in Table \ref{tab: SNR}. These results indicate that the cross-correlation between HI and CMB lensing could be a promising probe only if it can be precisely debiased. 

CMB lensing can trace all the dark matter distribution in the universe while the galaxy surveys can achieve tomographic mapping of the dark matter distribution. Different cross- and auto-power spectra formed by a set of three observables $\{\rm HI, \kappa, g\}$ can further improve the detectability of the faint HI signals and would be important probes to fundamental physics in late universe.\\

\begin{deluxetable*}{cccccc}
\tablecaption{The global SNR for each configuration. For \lz\, the SNRs with leakage corrections are slightly higher than the theoretical expectations. This is because we neglected the variance associated with the leakage correction terms. An exact computation of the leakage correction variance is beyond the scope of this work. \label{tab: SNR}}
\tablecolumns{6}
\tablenum{2}
\tablewidth{0pt}
\tablehead{
\colhead{HI regime} &
\colhead{Corr. channels\tablenotemark{a}} & 
\colhead{$n_s$\tablenotemark{b}} & 
\multicolumn{3}{c}{SNR}\\ 
\cline{4-6}
\colhead{} & \colhead{} & \colhead{} & 
\colhead{no corrected} & \colhead{corrected} & \colhead{simulated}
}
\startdata
\textbf{LOW-Z} &   & 3 & 1.3 & 5.0 & 4.7 \\
               & 6 & 4 & 1.3 & 5.0 & 4.7\\
               &   & 5 & 1.3 & 5.0 & 4.7\\
\cline{2-6}               
               &    & 3 & 1.3 & 33.0 & 31.7\\
               & 30 & 4 & 1.3 & 33.3 & 31.7\\
               &    & 5 & 1.3 & 34.2 & 31.7\\   
\hline               
\textbf{HIGH-Z}&   & 3 & 0.7 & 0.8 & 3.0 \\
               & 6 & 4 & 1.8 & 2.7 & 3.0\\
               &   & 5 & 1.0 & 1.6 & 3.0\\
\cline{2-6}                      
               &    & 3 & 5.2 & 21.2 & 47.6\\
               & 70 & 4 & 2.6 & 47.5 & 47.6\\
               &    & 5 & 1.8 & 21.2 & 47.6\\   
\enddata
\tablenotetext{a}{Number of HI channels correlated with the convergence field}
\tablenotetext{b}{Number of FastICA independent components}
\end{deluxetable*}

\section{Conclusions}
\label{Section: 8}
There have been several detections of HI and galaxy cross-correlations, however, the cross-correlations between HI and the CMB lensing would not only be complementary but robust cosmological probes since they are less contaminated by systematic issues. Moreover, CMB lensing can map all the LSS structures along the line of sight, directly tracing the dark matter distribution.

The HI IM analyses normally adopt blind foreground removal methods with minimal assumptions. Therefore, the foreground removal procedure inevitably removes the radial HI modes because there is degeneracy in long radial wavelengths between HI and foregrounds. The reduction of these long-wavelength modes has a negligible impact on the HI autocorrelations at two different frequencies. However, it can significantly suppress the cross-correlations between HI and CMB lensing. 

We generated correlated simulations for the HI and $\kappa$ fields and produced mock observations at different frequencies, taking into account the foreground contamination and instrumental noises. We verified that the cross-power spectra of the input simulations are consistent with the theoretical predictions. The blind foreground removal methods were applied to these mock observations, and reconstructed HI maps were cross-correlated with the simulated CMB lensing maps. To investigate different components of the cross correlations, we performed various tests for two scenarios, including a case when all HI channels are correlated with $\kappa$ and a case when a subset of HI channels are correlated with $\kappa$. The tests indicate that the cross-power spectra can be moderately suppressed if only a limited number of HI channels are correlated with the CMB lensing.  

We made forecasts for cross-correlations between the HI IM experiments including \lz\, and \hz\, scenarios, and the LSS tracers including CMB lensing and galaxy overdensity. The signal-to-noise ratios per multipole were estimated for the cases without and with suppression corrections. 

In addition, we theoretically modeled the leakage components, which were validated with simulations. The debiased cross-power spectra are consistent with theoretical expectations. Moreover, we found that better SNRs can be achieved at linear regions such as $\ell \sim 300$ after the leakage corrections. 

Future IM experiments will obtain multi-frequency HI datasets with low instrumental noises, and next-generation CMB experiments will produce a high-sensitivity lensing map via the quadratic estimator technique. Therefore, it is conceptually straightforward to cross-correlate the HI observations with the LSS tracers, such as the CMB lensing, to infer the underlying astrophysical and cosmological information. This work explores the detectability of the cross-power spectra and performs theoretical and numerical calculations to analyze details of the complex signals, paving the way for using HI-lensing cross-correlations as complementary cosmological probes.

\section*{Acknowledgments}
\begin{acknowledgments}
We are grateful for the helpful discussions with Xin Zhang, Yichao Li, David Bacon, and Jiajun Zhang. We also thank Dongdong Zhang for helping estimate the weak lensing noise. This work is supported by the USTC's starting grant. We acknowledge the use of \healpix~\citep{HEALpix}, \textsc{matplotlib}
\citep{matplotlib}, \textsc{numpy} \citep{numpy}, \textsc{scikit-learn} \citep{scikit-learn}, \textsc{CAMB} \citep{camb} and \textsc{NaMaster} \citep{namaster}. 
\end{acknowledgments}

\bibliography{biblio}{}
\bibliographystyle{aasjournal}
\end{document}